# How Software Engineers Engage with AI: A Pragmatic Workflow

Vahid Garousi -- Azerbaijan Technical University; Queen's University Belfast
Zafar Jafarov, Aytan Mövsümova, Atif Namazov -- Azerbaijan Technical University

**Abstract:**

Artificial Intelligence (AI) tools such as GitHub Copilot and ChatGPT are increasingly used in software engineering (SE) for tasks such as code, test, and documentation generation. However, engineers often face uncertainty about when to trust, refine, or discard AI-generated artifacts. We present a pragmatic workflow, complemented by a four-quadrant decision model, that formalizes how developers iteratively prompt, inspect, refine, and, when needed, fall back to manual work. The workflow and decision model were derived from a grey literature review and field observations across three industrial settings in Türkiye and Azerbaijan. Two real-world scenarios demonstrate the workflow's practical value, showing how engineers navigate key decision points when using AI. Our approach offers lightweight, structured guidance to support more deliberate and quality-aware use of AI tools in everyday SE tasks.

## 1 INTRODUCTION

AI-powered tools such as GitHub Copilot and ChatGPT are increasingly used in software engineering (SE) for almost any SE task that human software engineers have been doing for years, e.g., generating code, tests and documentation, inspection of SE artifacts, analyzing defect data and execution logs, etc. When AI is used for coding, one increasingly visible practice is often referred to as "vibe coding," which reflects exploratory, prompt-driven co-development of code together with AI tools[1-3]. Vibe coding represents only one mode of AI-assisted SE; other approaches include agentic AI systems, where autonomous or semi-autonomous agents plan, coordinate, and execute development tasks with varying degrees of human oversight.

While adoption of AI in SE is rising at an astonishing rate, concerns persist about the reliability, maintainability, and completeness of AI-generated artifacts [1, 2, 4]. As a leading tool vendor has noted, "*AI-generated code often undergoes less scrutiny than code we write ourselves, leading to an accountability crisis in software development*" [5]. Studies are finding that AI-generated outputs frequently require significant review and refinement by engineers before integration [2, 4].

Despite this rapid adoption, little is known about how engineers actually engage with the AI tools during daily tasks. Existing literature largely evaluates tool capabilities or code quality, but there is a need for structured models of developer behavior when iteratively prompting, inspecting, and deciding whether to trust and use AI-generated artifacts, or when to discard them.

Some recent efforts have proposed initial process models for AI-assisted SE, e.g., [6]. However, these models are simplistic—depicting a single-pass prompt-to-output workflow without iterative refinement, fallback, or validation. They also often lack explicit decision points or quality reasoning, overlooking the nuanced, decision-driven behavior engineers exhibit when using AI tools in SE. In contrast, the model presented in this work characterizes and formalizes decision points and trade-offs that developers inherently follow when using AI tools in SE. Two real scenarios from industry settings illustrate how the model is applied in practice and show its usefulness. The model is primarily descriptive in nature, capturing observed decision-making practices as they occur in real-world AI-assisted software engineering, rather than prescribing a normative process to be followed.

In the next section, we first describe our systematic approach for designing the model, followed by its presentation and example applications.

## 2 APPROACH FOR DESIGNING THE MODEL

To better understand, characterize and analyze how software engineers interact with AI tools in practice, approach for designing the process model utilized two complementary sources of insights:

- A "rapid" review [7] of practitioner-authored reports and online resources (grey literature), in which similar process models (workflows) for AI-assisted SE have been proposed
- Field observations in three industrial settings, where we shadowed engineers as they used tools such as GitHub Copilot and ChatGPT during real SE tasks.

The rapid-review and field observations were carried out using structured but lightweight procedures, inspired by established guidelines [7, 8]. We briefly discuss these two activities in the following. Further details, including the rapid-review protocol, its search process, planning and execution of the field observations, and data analysis steps, are fully documented in a supplementary technical report [9].

## Review of existing models of software engineers' interactions with AI

We conducted the rapid grey-literature review to capture how practitioners describe AI-assisted SE workflows. Following guidelines for rapid [7] and grey literature reviews [10], we searched Google and Google Image Search using terms such as "vibe coding workflow" and "AI-assisted software engineering process".

Practitioner-authored blog posts and shared workflow diagrams were screened for relevance, and recurring patterns were synthesized. The review resulted in a pool of eight practitioner-authored workflows (can be found in [9]), and provided a broad view of existing workflow models, which informed the design of our model.

Our synthesis of practitioner-authored workflows revealed common limitations: (1) linear structure without fallback or branching, (2) no explicit decision points or criteria for accept/refine/discard, (3) vague iteration guidance, and (4) no quality checkpoints before integration. These gaps directly informed our design of a combined workflow and decision guide for structured, quality-aware AI tool use. Full comparative analysis is provided in the supplementary technical document [9].

Also, a small but growing body of empirical academic research has begun to examine how software engineers interact with generative AI tools and how such interactions shape decision making. Ogenrwot and Businge [11] analyzed how developers use ChatGPT during pull-request reviews, showing that AI suggestions actively influence patch acceptance decisions, while final integration remains a human judgment shaped by context and trust considerations. Ulfsnes et al. [12] provided qualitative evidence that generative AI alters day-to-day development practices by reshaping interaction patterns, collaboration, and individual decision making, particularly in how developers balance AI assistance against peer consultation and established team norms. While these studies provide valuable empirical insights into AI-supported interactions and decisions, they have not offered explicit, structured models of the decision points through which engineers evaluate, accept, adapt, or reject AI-generated artifacts.

## Field observations on how software engineers use AI and how they make the relevant decisions

Our field observations aimed at gathering complementary insights into how engineers navigated the interplay between AI-generated outputs and their decision-making w.r.t. the outputs that they get from AI, i.e., evaluating whether to accept, refine, or discard AI outputs. We recorded the sequence of actions—prompt creation, AI responses, inspection, refinement, and fallback to manual work—focusing particularly on decision points where engineers made the above decisions.

The participating teams, based in two large software companies in Türkiye and one in Azerbaijan, had already integrated AI tools into their daily practices. The observed teams operated in web-based enterprise software development, including backend service development and web application testing, with roles spanning backend developers and QA engineers.

Across these settings, we consistently observed that teams lacked a shared, structured approach for deciding when and how to use AI effectively. Engineers and team leads expressed a need for a model that formalizes the cognitive and procedural flow of AI-assisted development and offers guidance for reasoning about trade-offs between effort saved and artifact quality.

Across the three sites, we observed 11 unique engineers, each for approximately 90 to 120 minutes, depending on their task. Observation sessions were balanced across the companies: four participants at Site A (Türkiye), four at Site B (Türkiye), and three at Site C (Azerbaijan). Sessions involved non-intrusive shadowing of real development tasks, focusing on engineers' engagement with tools such as GitHub Copilot and ChatGPT. Prompts, outputs, and decision-making behaviors were documented in situ and followed by brief post-session reflections. Data collection took place in early Summer 2025, during which participants used contemporary versions of widely adopted AI tools such as GitHub Copilot and ChatGPT available at that time.

Insights gathered from the field observations, combined with the synthesized findings from the grey-literature review, provided the foundation for the AI-SE workflow and decision guide introduced in the following section.

# 3 The AI-SE activity process model

Figure 1 presents our proposed workflow and decision model. The workflow captures the iterative, decision-driven interactions observed when engineers use AI tools such as GitHub Copilot and ChatGPT, making these interactions explicit without prescribing a specific way of working.

To improve readability and traceability, the workflow in Figure 1a explicitly marks six key decision points (D1–D6), which structure how engineers reason about AI use, quality, and fallback throughout an AI-assisted software engineering activity. We reference these decision points explicitly below.

The workflow begins when an engineer engages in a task—e.g., coding, test creation, or documentation. The first decision point (D1 in Figure 1a) determines the desired artifact type: whether the engineer seeks a boilerplate (template) (e.g., a class skeleton) or a nearly complete artifact (e.g., a complex SQL query). Engineers typically treat these two artifact types differently: boilerplate tasks (e.g., class skeletons) are well-suited for quick AI generation with minimal inspection, whereas nearly-complete artifacts (e.g., complex SQL queries or detailed test scripts) require richer prompts, more iterations, and closer human oversight. Introducing this decision point early in the workflow clarifies that prompt detail, review effort, and fallback likelihood vary significantly by artifact type, making the model more realistic, actionable, and aligned with real-world developer behavior.

For boilerplate generation, the process typically involves writing a short prompt and quickly evaluating the output. Decision point D3 captures this initial quality judgment for boilerplate artifacts, where engineers decide whether the generated template is sufficient to proceed or requires further refinement. At decision point D2, the engineer decides whether to continue using AI assistance for the current activity or to abandon AI early and proceed manually.

For nearly complete artifacts, prompts are (should be) more detailed and contextualized. Outputs undergo explicit inspection, which may include manual review / analysis. Engineers decide whether to accept, refine, or discard the artifact, with fallback to manual creation if quality is insufficient. Decision point D4 represents the core quality assessment step for near-complete artifacts, where human inspection—manual or tool-assisted—determines whether the AI-generated output meets acceptable quality thresholds. The loop-and-fallback structure reflects the trial-and-error refinement that we have consistently observed in practice, and also reported in many other studies [13, 14].

At decision point D5, engineers explicitly decide whether further prompt refinement is worthwhile or whether continued AI use no longer justifies the effort. Decision point D6 distinguishes between total abandonment of AI-generated artifacts and partial reuse, where fragments of AI output are retained and manually repaired or adapted.

The four-quadrant decision model in Figure 1b is labeled Q1–Q4 and is invoked at the evaluation points in the workflow (notably D3 and D4) to guide accept, refine, or discard decisions. The decision model complements the workflow by visualizing trade-offs between effort saved and expected artifact quality. Locating an output within one of the four quadrants helps engineers decide to accept, refine, or discard it. For example:

- Quadrant Q1 (high quality, high effort saved) corresponds to outputs that can typically be accepted with minimal edits.
- Quadrant Q2 (low quality, high effort saved) indicates outputs that appear productive but require careful inspection and often refinement or rejection.
- Quadrant Q3 (low quality, low effort saved) represents outputs that provide little value and are best discarded..
- Quadrant Q4 (high quality, low effort saved) captures cases where manual development may have been equally or more efficient.

Assignment of an AI-generated artifact to one of the four quadrants is intentionally heuristic and judgment-based. Engineers qualitatively assess (i) the expected quality of the artifact relative to task requirements (e.g., correctness, completeness, maintainability) and (ii) the net effort saved compared to manual development, considering prompt creation, inspection, refinement, and rework. The model does not prescribe numerical thresholds, reflecting how such judgments are made pragmatically in real-world practice. By placing an AI-generated artifact within the four-quadrant space, engineers can decide how to proceed. Outputs in the top-right quadrant (high quality, high effort saved) can typically be accepted with minimal edits. Top-left outputs warrant thorough inspection before use. Bottom-right outputs are of good quality but may not justify the AI effort. Bottom-left outputs provide little value and are best discarded.

Movement between quadrants represents a change in the engineer's qualitative assessment rather than crossing a fixed threshold. For example, deeper inspection may reveal hidden defects that lower perceived quality, or repeated refinements may reduce perceived effort savings, causing an artifact to shift from one quadrant to another as judgment evolves.

Such reassessments typically occur at decision points D3 and D4 in the workflow (Figure 1a), where engineers re-evaluate quality and effort before deciding to accept, refine, or abandon AI-generated artifacts.

Together, the workflow and decision model provide a structured yet lightweight approach for navigating AI-assisted SE tasks. They externalize the reasoning process engineers typically follow informally, supporting deliberate and quality-aware use of AI outputs.

While AI tools can be helpful for routine or low-complexity tasks, their effectiveness decreases for complex or context-dependent activities. We observed—and practitioner reports confirm [15-17]—that developers often abandon AI after repeated low-quality results, reverting to manual development. A pragmatic strategy is to decompose complex tasks into smaller sub-tasks, applying AI selectively where it performs well and relying on human expertise for integration and final quality assurance. These insights reinforce the need for a workflow with explicit decision points and fallback paths, as captured in our proposed model.

## 4 TWO REAL-WORLD SCENARIOS ILLUSTRATING USAGE OF THE WORKFLOW AND DECISION MODEL

We have actively applied the workflow and decision model in the three industrial settings for several months. More than 12 software engineers have actively used them in their routine SE tasks. For brevity, we present in the following two scenarios that serve as representative examples, randomly sampled from many usages of the workflow and decision model in the industrial settings. In each case, a software engineer who was working on a SE task, as part of her/his routine work, was involved, wo was provided with the workflow and decision model, and was advised to use them, and later to assess their usefulness.

As we see below, the two scenarios show two contrasting paths through our workflow: (1) successful AI use for a simple boilerplate task; and (2) fallback to manual development after unsuccessful usage of AI for a complex SE task. Although both scenarios below involve code generation, they represent distinct software engineering activities—API development and test automation—each with different quality concerns, artifact structures, and risk profiles.

### Scenario 1: Generating REST Controller Boilerplate Code

At a large Turkish enterprise, a backend developer used GitHub Copilot to generate boilerplate Spring Boot REST controller classes.

- **Step 1 of the workflow – Task initiation:** The developer identified the task as boilerplate code generation.
- **Step 2 – Prompt creation:** A short natural-language prompt described the required REST controller endpoints. The actual prompt templates used by engineers in these scenarios, created using prompt-engineering best practices, are provide in Table 1.
- **Step 3 – AI output and inspection:** Copilot generated a controller class with correct structure, imports, and annotations. A quick manual inspection confirmed syntactic correctness and alignment with coding conventions.
- **Step 4 – Decision point:** The artifact was accepted without edits, and integrated into the project's code-base.

**Decision model interpretation:** The output was placed in the top-right quadrant (high quality, high effort saved). The workflow followed the fast-loop path, where AI output is immediately useful with minimal review.

### Scenario 2: Creating Selenium Test Scripts

At a small startup in Azerbaijan, a QA engineer used ChatGPT to generate Selenium-based UI test scripts.

- **Step 1 of the workflow – Task initiation:** The engineer aimed to generate a nearly complete UI test suite.
- **Step 2 – Prompt creation:** A detailed prompt was developed, telling AI the type of test cases needed, browser requirements, and sample page elements (Table 1).
- **Step 3 – AI output and inspection:** ChatGPT produced a plausible-looking test script. On closer inspection, brittle CSS selectors, redundant waits, and incorrect assertions were found.
- **Step 4 – Decision point:** The engineer refined the prompt and re-generated the script twice, but the quality remained unsatisfactory.
- **Step 5 – Fallback:** The engineer abandoned AI assistance and manually rewrote the test scripts.

**Decision model interpretation:** The final outputs fell into the **top**-left quadrant (low quality, high effort saved). Despite apparent productivity gains, risks of fragile and incorrect tests outweighed the benefits. The workflow's fallback branch was followed, illustrating how developers critically evaluate AI outputs before deciding to abandon them.

**Table 1- AI prompt templates used in the two scenarios**

| Prompt template for generating boilerplate REST controllers (scenario 1): |
| --- |
| Generate a Spring Boot REST controller class in Java named OrderController that exposes the following endpoints:<br>- GET /orders<br>- POST /orders<br>- GET /orders/{id}<br>Include basic request/response mappings and use appropriate annotations (@RestController, @RequestMapping, @GetMapping, @PostMapping). Do not include any business logic, only method stubs. |
| Prompt template for generating Selenium test scripts (scenario 2): |
| Generate Selenium test scripts in Java for the Chrome browser to test the login functionality of a web application. The test steps are:<br>1. Open the login page at https://example.com/login<br>2. Enter a username and password<br>3. Click the "Login" button<br>4. Verify that the page title is "Dashboard"<br>5. Verify that the welcome message contains the logged-in username.<br>Use CSS selectors for locating elements. Include meaningful assertions for each step. Avoid hardcoded sleep times; use appropriate waits instead. |

### Findings from the illustrative scenarios

These scenarios are not isolated anecdotes; they represent patterns observed repeatedly across our broader field study. Engineers often successfully use AI for routine tasks but abandon AI after repeated low-quality results for complex, context-dependent tasks, a challenge also reported by other practitioners [15-17]. Such cases align with broader observations that AI tends to struggle with SE tasks which have complex logic and deep contextual dependencies [18].

By walking through each workflow step and applying the four-quadrant decision model, these scenarios demonstrate how engineers actively evaluate AI outputs, refining or discarding them when necessary, rather than passively accepting generated artifacts.

## 5 Discussion and Concluding Remarks

Our work introduces an AI-SE workflow and four-quadrant decision model that formalize how engineers engage with AI tools when generating software artifacts. Unlike existing linear "vibe-coding" workflows, our model incorporates explicit decision points, fallback strategies, and a visual guide for balancing effort saved and artifact quality. The model is based on a rapid grey-literature review of practitioner-authored sources and direct field observations in three industrial settings.

### Practical Use and Value

The workflow and decision model are deliberately lightweight so that developers can apply them without disrupting day-to-day activities. They externalize reasoning processes that many engineers already follow informally—iterative prompting, output inspection, and fallback to manual work when AI results are inadequate. By making these steps explicit, the model supports decision-making and risk awareness, helping teams avoid pitfalls such as automation bias or over-reliance on superficially plausible outputs.

After continuous usage of the workflow and decision model by 12 software engineers in the three companies for more than two months, we invited them to complete a short online anonymous questionnaire with rubric-style ratings on the clarity, applicability, and usefulness of emerging workflow concepts (5-point Likert scales). Design details of the questionnaire can be found in supplementary technical report [9].

Participants gave average ratings of 4.5/5 for the workflow's clarity, 4.2/5 for its perceived applicability, and 4.3/5 for its perceived usefulness. Several noted that the four-quadrant diagram helped them explain "why" they accepted or rejected AI outputs, making decision-making more explicit. These benefits should be interpreted as practitioner-perceived and observation-based, rather than as outcomes of a controlled experimental evaluation.

We furthermore emphasize that these observations describe how engineers currently reason about AI-assisted work, rather than prescribing how such work should be conducted.

These observations complement, but also differ in emphasis from the preliminary experimental results of Leotta et al. [19], who found that AI-generated test scripts generally reduce development time compared to fully manual approaches, provided that testers are skilled and able to repair AI output. While both studies identify similar quality and correctness issues in AI-generated tests, our work explicitly models the situated decision-making process through which engineers may still rationally prefer partial reuse or a full return to manual development when the perceived risks outweigh the benefits. This difference reflects our focus on micro-level decision episodes during AI use, rather than aggregate time savings across tasks.

### Contribution Beyond Existing Models

While our model is specific to AI-assisted SE, it relates to and extends broader concepts in software process modeling and decision-support. Traditional SE models (e.g., waterfall, Agile workflows) prescribe task sequences but rarely account for real-time human–AI interaction or iterative decision cycles during artifact creation. Similarly, cognitive decision-making frameworks in SE [20] often focus on architectural or team-level trade-offs, not the micro-level decisions developers make while prompting and reviewing AI outputs. By centering on micro-level, artifact-oriented decision points during real-time prompting and artifact inspection, our model fills a gap between abstract decision theory and actual developer interactions with AI.

Our review showed that the existing AI-assisted SE workflows lack structured decision logic. Our contribution consolidates practitioner insights into a coherent, empirically informed model that integrates both the activity flow and a decision guide, making it more actionable than previous informal workflows. By walking through real scenarios step-by-step, we show how engineers traverse the workflow, apply decision points, and ultimately choose whether to accept, refine, or discard AI outputs.

### Closing Perspective

AI tools such as GitHub Copilot and ChatGPT are rapidly evolving, yet their integration into real-world SE practices remains non-trivial. By capturing how engineers iteratively prompt, inspect, and decide whether to trust AI-generated artifacts, our workflow and decision model provide structured guidance for quality-aware use of AI in SE.

Rather than promoting uncritical adoption, our work highlights the importance of deliberate human oversight, helping practitioners reason about trade-offs between effort saved and artifact quality. In this way, the model serves as both a reflection of current best practice and a practical aid for engineers navigating the evolving landscape of AI-assisted SE.

### Limitations and Future Work

Like any conceptual model, ours assumes that engineers apply critical thinking when reviewing AI outputs. However, real-world factors such as cognitive fatigue, automation bias, or organizational pressure to "move fast" may lead to suboptimal decisions that deviate from the model's guidance.

Moreover, although our field observations included multiple engineers across three organizations, these were primarily early adopters in Türkiye and Azerbaijan. The applicability of the model across different organizational cultures, AI maturity levels, and SE roles (e.g., architects, analysts) remains an open question. Team-based dynamics, such as shared prompting strategies or code review practices, were also beyond the scope of this study.

A further limitation is that the illustrative scenarios focus on code-centric tasks and do not yet cover upstream software engineering activities such as requirements specification or system design (e.g., UML modeling). Addressing these tasks would require additional, dedicated industrial engagements, which were beyond the scope of the current study. We explicitly plan to extend the model and its empirical grounding to design- and requirements-oriented scenarios as part of an ongoing larger-scale follow-up study.

An important direction for future work is to explore decision-support mechanisms (e.g., lightweight rules, metrics, or tool support) to assist engineers in making the qualitative quadrant judgments (quality versus effort, shown in Figure 1b) more systematically, while preserving practitioner control.

Another limitation is that we have not yet conducted a controlled empirical evaluation to measure the impact of the model on developers' productivity, decision quality, or satisfaction. Future research should include experimental studies and broader surveys to evaluate these aspects. Directions for future work include:

- Quantitative validation of the model's impact on software quality, productivity, and engineer satisfaction.
- Adapting the model for various SE roles (e.g., testers, documentation writers, architects).
- Investigating its use in collaborative settings such as pair programming and AI-assisted code review.
- Integrating the model into development environments (e.g., via IDE plugins) to study adoption and usage patterns in situ.

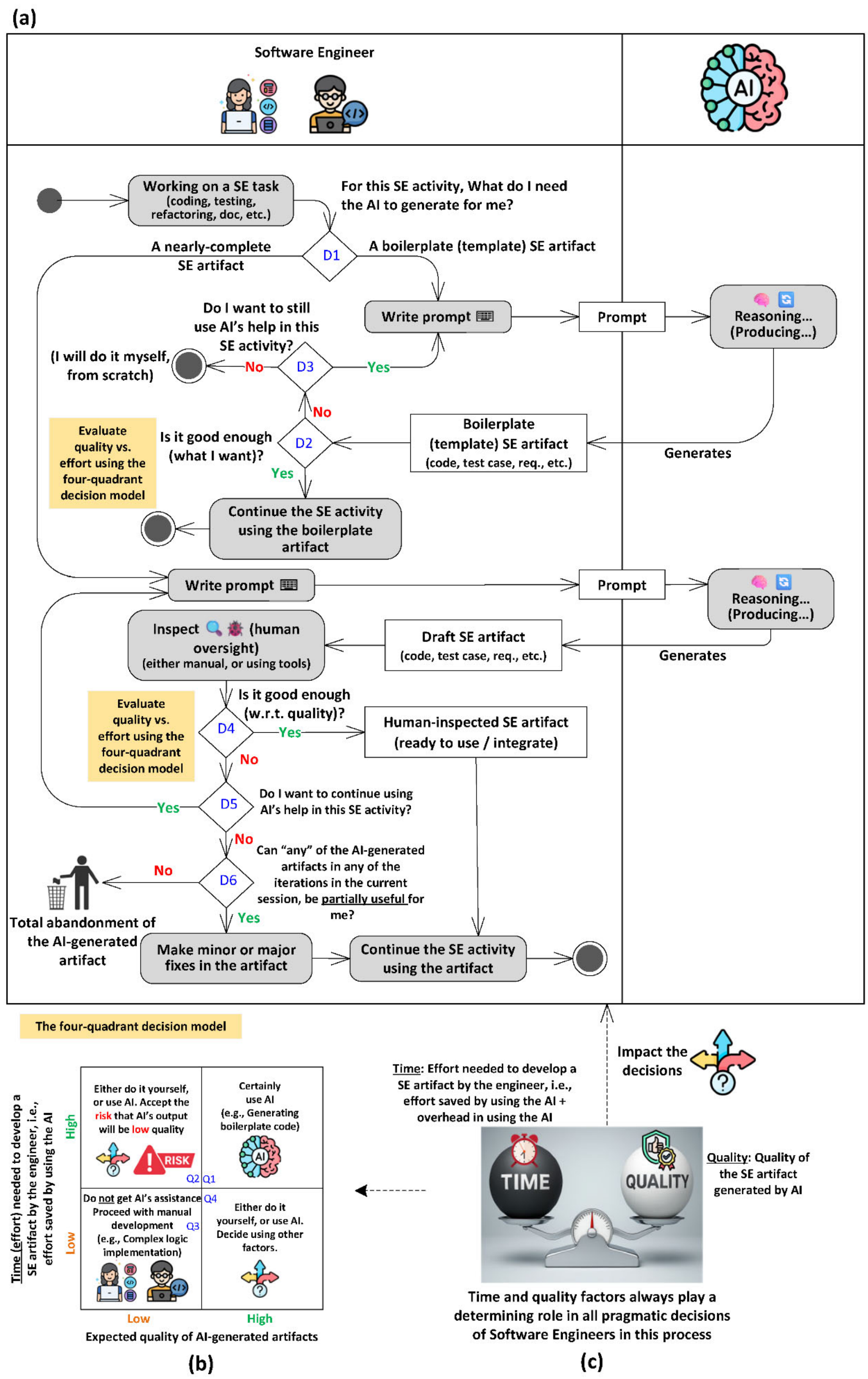

**Figure 1-A process model of AI-assisted SE, including a four-quadrant decision model**